\documentclass[prl,twocolumn,letterpaper,10pt]{revtex4-1}
\usepackage{graphicx}
\usepackage[final,dvips]{epsfig}
\usepackage{color}
\usepackage{dcolumn}  
\usepackage{bm}       
\usepackage{array}    
\usepackage{amssymb,amsmath}
\usepackage{appendix}
\usepackage{hyperref}

\emergencystretch=\maxdimen
\usepackage{natbib}

\definecolor{darkgreen}{rgb}{0,0.7,0}
\hypersetup{
	colorlinks=true,
	linkcolor=blue,
	citecolor=darkgreen
}

\newcommand{\be}{\begin{equation}}
\newcommand{\ee}{\end{equation}}
\newcommand{\ba}{\begin{eqnarray}}
\newcommand{\ea}{\end{eqnarray}}

\newcommand{\upa}{\uparrow}
\newcommand{\dna}{\downarrow}

\newcommand{\COMMENTED}[1]{}

\newcommand{\ca}[2]{c_{#1 #2}^\dagger}
\newcommand{\de}[2]{c_{#1 #2}^{\phantom{\dagger}}}

\newcommand{\ob}[1]{{\langle #1\rangle}}

\newcommand{\bfr}{\mathbf{r}}

\newcommand{\reff}[1]{{\color{blue} Fig.\ \ref{fig:#1}}}
\newcommand{\reffa}[1]{{\color{blue} Fig.\ \ref{fig:#1}(a)}}
\newcommand{\reffb}[1]{{\color{blue} Fig.\ \ref{fig:#1}(b)}}
\newcommand{\reffc}[1]{{\color{blue} Fig.\ \ref{fig:#1}(c)}}
\newcommand{\reffd}[1]{{\color{blue} Fig.\ \ref{fig:#1}(d)}}

\newcommand{\reffld}[1]{{\color{blue} Figure\ \ref{fig:#1}(d)}}
\newcommand{\refq}[1]{Eq.\ (\ref{eq:#1})}
\newcommand{\refqb}[1]{(\ref{eq:#1})}
\newcommand{\myparagraph}[1]{{\it #1} -- }

\begin{document}

\title{Discriminating antiferromagnetic signatures in ultracold fermions\\ by tunable geometric frustration}

\author{Chia-Chen Chang}
\affiliation{Department of Physics, University of California, Davis, California 95616 USA}
\author{Richard T. Scalettar}
\affiliation{Department of Physics, University of California, Davis, California 95616 USA}
\author{Elena V. Gorelik}
\affiliation{Institute of Physics, Johannes Gutenberg University, Mainz, Germany}
\author{Nils Bl\"umer}
\affiliation{Institute of Physics, Johannes Gutenberg University, Mainz, Germany}

\begin{abstract}
Recently, it has become possible to tune optical lattices continuously between square and triangular 
geometries. We compute thermodynamics and spin correlations in the corresponding Hubbard model using 
determinant quantum Monte Carlo and show that the frustration effects induced by the variable hopping 
terms can be clearly separated from concomitant bandwidth changes by a proper rescaling of the interaction. 
An enhancement of the double occupancy by geometric frustration signals the destruction of nontrivial 
antiferromagnetic correlations at weak coupling and entropy $s\lesssim \ln(2)$ (and restores Pomeranchuk 
cooling at strong frustration),  paving the way to the long-sought experimental detection of 
antiferromagnetism in ultracold fermions on optical lattices. 
\end{abstract}

\maketitle

The comparison of the physics of
antiferromagnetism on bipartite and frustrated lattices, and the
interpolation between them, is a fascinating topic already at the classical level.  
In the Ising model on a square lattice with antiferromagnetic (AF) nearest-neighbor
exchange $J$ and an additional AF coupling $J'$ along one of the
diagonals, long-range AF order appears below $T_{\text{N}}= 2(J-J')/\ln(2)$ at $J'<J$
\cite{Stephenson:1964,*Stephenson:1970a,*Stephenson:1970b,Eggarter:1975}.  
The physics is more complex than this, however: at $T>T_{\text{N}}$, AF order persists 
at intermediate ranges up to a `disorder line' $T_d(J,J')$, above which
it  becomes incommensurate, i.e.,
the peak in the structure factor moves away from the AF wave vector
$\mathbf Q=(\pi,\pi)$.

Quantum physics can be introduced into such a classical model via a
transverse magnetic field $B_{\perp}$.  In the absence of frustration,
these quantum fluctuations compete with magnetic order and drive an
AF to paramagnetic (PM) phase transition.  In contrast,
$B_{\perp}$ can act to {\it induce} order when starting from the
classical model on a triangular lattice by lifting the ground state
degeneracy.  A rich set of phases results from the interplay of quantum
and thermal fluctuations, including two distinct ordered
phases \cite{Isakov:2003}.
There is considerable experimental interest in
realizing such frustrated quantum models in cold atomic gases and in the
observation of these effects \cite{Edwards:2010,Britton:2012}.  

In the present paper, we will examine related frustration physics in the context of
an itinerant model of magnetism, the $t$-$t'$ Hubbard Hamiltonian,
\ba
  H &=& -t\sum_{\ob{i,j},\sigma} \left( \ca{i}{\sigma}\de{j}{\sigma} + \mbox{h.c.} \right) 
   -t'\sum_{\ob{\ob{i,j}},\sigma} \left( \ca{i}{\sigma}\de{j}{\sigma} + \mbox{h.c.} \right) \nonumber\\
    & & + U \sum_{i} n_{i\upa} n_{i\dna}
    - \mu \sum_{i,\sigma} n_{i\sigma},
\label{eq:ham}
\ea
where $t$ is the nearest-neighbor  hopping amplitude on a square lattice and 
$t'$ is the next-nearest-neighbor  hopping along one of the diagonals, as depicted 
in the inset of \reffb{U4}
\footnote{While the square and triangular lattices shown in the inset of \reffb{U4} are topologically equivalent, our notation of lattice vectors (e.g. for correlation functions) is specific to the square geometry, with unit lattice spacing.}%
; $U>0$ is the repulsive onsite interaction. The chemical potential $\mu$ 
is tuned so that the system stays at half filling, unless otherwise noted.

\begin{figure*} 
  \includegraphics[width=\textwidth]{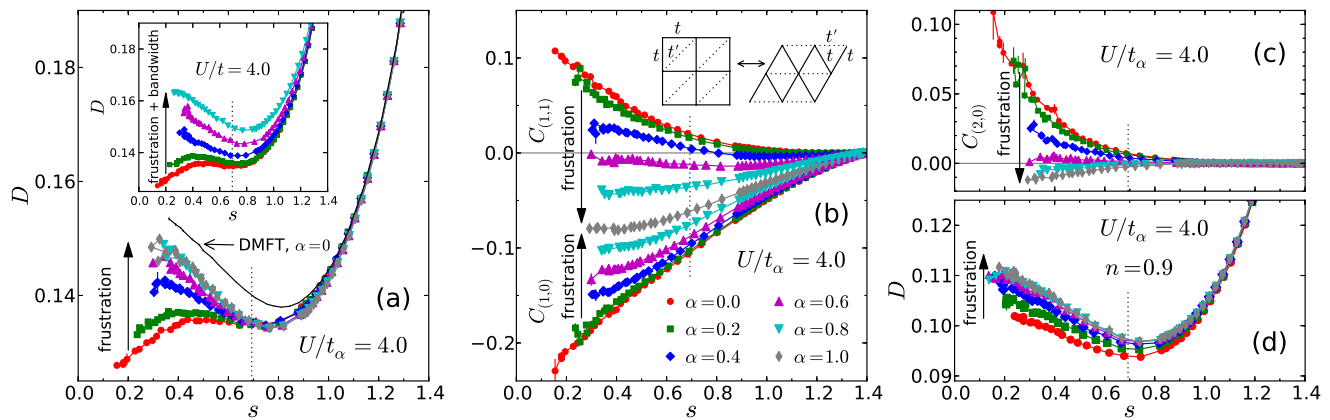}
  \caption{(Color Online) (a) Double occupancy $D$ versus entropy $s$ at weak coupling and variable 
  frustration $\alpha=t'/t$, with scaled interaction
  $U = 4 t_\alpha = 4 t \sqrt{1+\alpha^2/2}$; solid line: DMFT estimate of $D(s)$ at $\alpha=0$. 
  Inset: At unscaled $U = 4t$, $D(s)$ do not collapse.
  (b) Short-range spin correlation functions $C_{(1,0)}$ and $C_{(1,1)}$ [equivalent in triangular 
  limit $\alpha=1$] at $U= 4 t_\alpha$. Inset: Square lattice geometry used in simulations and 
  topologically equivalent anisotropic triangular lattice.
  (c) Longer-range spin correlation $C_{(2,0)}$. 
  (d) $D(s)$  at density $n=0.9$ and frustration $\alpha$
  at $U/t_\alpha=4$. Vertical dotted lines indicate $s^*\equiv\ln(2)$.}
  \label{fig:U4}
\end{figure*}

Our purpose is two-fold.  The first and primary motivation is to provide guidance for
the next generation of experiments of quantum magnetism on ultracold fermions on optical lattices
and to establish precise numerical reference results.
Cold-atom experiments have demonstrated 
the Mott metal-insulator transition \cite{Jordens:2008,Schneider:2008}.  However, the
observation of quantum magnetism has proven much more challenging, owing
to the low temperature scales required, with much of the success limited
to classical and bosonic systems \cite{Trotzky:2008,Simon:2011,Struck:2011}.  
Very recent experiments have realized tunable lattice geometries for cold fermions \cite{Greif:2012} 
and bosons \cite{Jo:2012} with the goal of emulating magnetic or superfluid phases in many-body 
systems. In particular, nearest-neighbor AF correlations of fermionic atoms have been observed
on dimerized and anisotropic geometries \cite{Greif:2012} 
with planned extensions to
honeycomb and triangular lattice geometries \cite{Greif:2012,Meng:2010}. 
It is with this latter objective in mind, with its attendant promise of searching 
for spin-liquid and other exotic physics, that we simulate the $t$-$t'$ Hubbard model.  

A second goal is to expand our understanding of itinerant antiferromagnetism in
frustrated geometries.
Like the next-nearest-neighbor exchange $J'$ on a
square lattice, the hopping $t'$ induces an AF 
superexchange interaction which can be expected to push the ordering wave
vector away from $\mathbf Q=(\pi,\pi)$.  Thus the $t$-$t'$ Hubbard Hamiltonian
is a natural generalization of spin models capturing the interplay of 
quantum and thermal fluctuations, and frustrating interactions.

We solve the Hamiltonian \refqb{ham} using determinant 
quantum Monte Carlo (DQMC) \cite{Blankenbecler:1981,White:1989}. The method is 
exact, apart from statistical errors which can be reduced by increasing the
sampling time, and Trotter errors associated with the discretization
of the inverse temperature $\beta\equiv 1/(k_{\text B}T)=\Lambda\Delta\tau$ 
(here, $\Lambda$ is an integer and $k_{\text B}$ is the Boltzmann constant, 
set to unity in the following), which can be eliminated by
extrapolation to the $\Delta \tau=0$ limit \cite{Gorelik:2012,Rost:2013}.  
Unless one is protected by particle-hole symmetry [as, e.g., for the Hamiltonian 
\refqb{ham} at $t'=0$ and half filling], the fermionic sign problem 
\cite{Loh:1990,Loh:2005,Troyer:2005} limits the accessible temperatures.

To interpolate between square and triangular lattices, we vary $t'$ 
in the range $[0,t]$ (as is now possible in cold-atom 
experiments \cite{Greif:2012}) and use $\alpha\equiv t'/t$ as a dimensionless scale.
Note that the addition of $t'$ increases the coordination number from $Z=4$ in the 
square lattice case ($\alpha=0$) to $Z=6$ in the triangular case ($\alpha=1$), and, 
correspondingly, the noninteracting bandwidth. 
This effect may be quantified using the new energy scale $t_\alpha$:
\begin{equation}\label{eq:scaling}
	t_\alpha=t\sqrt{Z_\alpha/Z_0}\,; \quad 
	Z_\alpha = 4 + 2\alpha^2.
\end{equation}
We will mostly use the dimensionless entropy $s=S/(k_{\text{B}}N)$ per particle 
as a thermal parameter \cite{Gorelik:2012}, as appropriate for (approximately 
adiabatic) cold-atom experiments.
In the following, we present DQMC data obtained for $8\times 8$ clusters at 
$\Delta\tau\, t_\alpha \leq 0.04$; we have verified (using additional simulations 
at $0.05 \le \Delta\tau\, t_\alpha \leq 0.1$) that the resulting Trotter errors
are insignificant and (using cluster sizes up to $16\times 16$) that finite-size
effects do not impact any of the conclusions 
(see Supplement). 


\myparagraph{Results at weak coupling} 
As is well-known, the additional hopping terms at $\alpha>0$ frustrate AF correlations 
by introducing superexchange between sites that would have the same local spin orientation 
in a perfect N\'eel state. Moreover, the increase in noninteracting bandwidth
weakens the relative impact of a fixed interaction 
$U$, i.e., makes the system less correlated. This is clearly seen in the inset of \reffa{U4} 
at $U/t=4$: increasing $\alpha$ shifts the double occupancy 
$D=\ob{n_{\bfr\upa}n_{\bfr\dna}}$ towards the uncorrelated limit $\ob{n_{\bfr\upa}}\ob{n_{\bfr\dna}}=1/4$,
regardless of entropy $s$; similar effects are observed also at stronger coupling (not shown). 
By scaling $U$ proportionally
to $t_\alpha$, these bandwidth effects can be eliminated,
as demonstrated in the main panel of \reffa{U4}. The DQMC estimates of $D(s)$ are 
here seen to collapse in the regime $s>s^*\equiv\ln(2)$
\footnote{Note that the entropy $s^*$ is not critical in any sense, not even at the mean-field level. 
In contrast, e.g., to the notation in \cite{Gorelik:2012}, it here just denotes a coherence scale.}.
Consequently, the remaining effect of $\alpha$, a strong enhancement of $D$ at $s<s^*$
must be associated with the suppression of (short-ranged) antiferromagnetism by geometric frustration. 
It is remarkable that this AF signature appears so sharply below $s^*$, even though the 
short-range spin correlation functions, shown in \reffb{U4}, vary smoothly as a function of $s$ 
(at fixed $\alpha$), with no particular features at $s\approx s^*$, and are sensitive to $\alpha$ 
at each $s$. Even in longer-range spin correlations, e.g. $C_{(2,0)}(s)$ depicted in \reffc{U4}, 
the AF signatures are less sharp than in $D(s)$.

Let us discuss the underlying physics in more detail. When charge fluctuations are
strongly suppressed at low temperatures,
a nonmagnetic state at half filling would be characterized by a random configuration with either a spin up 
or a spin down electron at each site, i.e., with $D=0$ and $s=s^*$. Higher entropies can only arise 
due to charge excitations, which ultimately drive $D\to 1/4$ and $s\to \ln(4)$ for $T\to\infty$ at all $U>0$. 
Lower entropies can be reached either by spin order or 
by long-range coherence of the 
charge quasiparticles, i.e., Fermi liquid physics. The latter effect is captured by ``paramagnetic'' 
dynamical mean-field theory (DMFT), which neglects nonlocal correlations and completely suppresses any spin 
order. At $s>s^*$, the DMFT estimate of $D(s)$ for the square lattice
is seen in \reffa{U4} 
(solid line) to converge to the $\alpha$-independent DQMC results, with increasing positive slope.
At low $s$,  $D$ is strongly enhanced, within DMFT, by Fermi liquid physics, giving rise to a negative slope 
$dD(s)/ds$ (or $dD(T)/dT$) 
in the regime $s<s^*$. 
This ``Pomeranchuk effect'' has been suggested as a tool for adiabatic cooling of cold atoms \cite{Werner:2005}.

However, such a negative slope is hardly seen in the DQMC data for the square lattice 
[red circles in \reffa{U4}]. Instead, $D(s)$ essentially forms a plateau in the range $0.4\lesssim s \lesssim 0.8$, 
and decays further at $s\lesssim 0.4$ when finite-range AF order develops \cite{Gorelik:preprint,Paiva:2010}. 
This deviation from the nonmagnetic DMFT prediction is caused by strong AF correlations, which destroy the 
charge coherence instrumental to the Fermi liquid enhancement of $D$. 
Geometric frustration at $\alpha>0$ should reduce these deviations by suppressing AF correlations. This is exactly 
what is observed in \reffa{U4}: with increasing $\alpha$, the Fermi liquid enhancement of $D$ is gradually restored. 
Note that this restoration is not complete, as AF correlations remain even in the triangular case, for which 120-degree 
order is expected at large $U$ \cite{Yoshioka:2009}.

\begin{figure} 
	\includegraphics[width=\columnwidth]{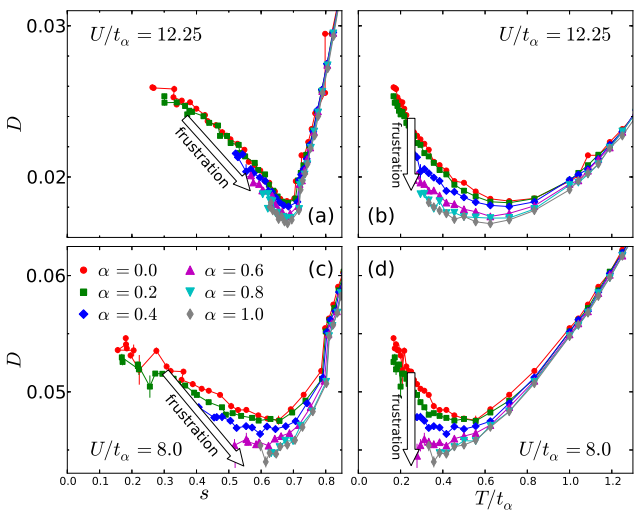}
	\caption{(Color online) Double occupancy $D$ versus entropy $s$ (left column) and temperature 
	$T/t_\alpha$ (right column) at strong coupling $U = 12.25\, t_\alpha$ (top) and $U = 8\, t_\alpha$ (bottom).
	}
	\label{fig:D}
\end{figure}

\reffld{U4} shows $D(s)$ measured at density $n=0.9$ and $U=4\,t_\alpha$. 
While the impact of $\alpha$ is qualitatively similar to that at $n=1$ [shown in \reffa{U4}], it is smaller,
especially in the range $\alpha\gtrsim 0.4$.  
This suggests that in cold-atom experiments, contributions from doped regions at the edge of the trap to measurements of $D$ averaging over the whole system \cite{Jordens:2008,Esslinger:2010} 
will hardly dilute the AF-specific signature predicted above from the analysis of the half-filled core. Moreover, the shell contributions are also arising from remnant AF correlations (which are quickly suppressed for $\alpha\gtrsim 0.4$ at $n=0.9$) and not from unrelated physics (such as Fermi liquid effects in the shell surrounding a Mott core \cite{Gorelik:2010,Khatami:2011}).


\myparagraph{Results at strong coupling}
At the strong interaction $U=12.25\,t_\alpha$, Fermi liquid effects are no longer relevant. Instead, $D$ is 
enhanced at low $s$, or $T$,  by AF correlations in the unfrustrated case \cite{Gorelik:2010,Gorelik:2012}, 
as seen in \reffa{D} and \reffb{D}, respectively (red circles).
\reffb{D} shows that geometric frustration has a drastic impact at fixed temperature: the low-$T$ enhancement 
of $D$ is almost completely eliminated (as illustrated by the arrow at $T/t_\alpha=0.25$) in the triangular 
limit $\alpha\to 1$, which is easily understood as the result of a strong suppression of AF correlations. 
In contrast, almost no impact of $\alpha$ is seen on the shape of the curves $D(s)$ in \reffa{D},  
due to a cancellation effect \footnote{By cancellation effect we mean, in general, a simultaneous shift by the 
frustration $\alpha$ of the entropy $s$ and the observable $O$ (at each fixed temperature) such that $O(s)$ 
remains largely unchanged.}: 
The frustration changes $D$ and $s$ (cf.\ \reff{entropy}) simultaneously, as indicated by the arrow in 
\reffa{D} for fixed $T/t_\alpha=0.25$. The net effect 
is just a shift on the same curve $D(s)$.
At $U=8\,t_\alpha$, this mechanism is only partially effective. Thus, the strong frustration effects visible 
in \reffd{D} at constant $T$ survive partially also at constant $s$, as seen in \reffc{D}.

Let us now turn to spin correlation functions at $U= 8\,t_\alpha$, depicted in \reff{corr}.
\begin{figure} 
	\includegraphics[width=\columnwidth]{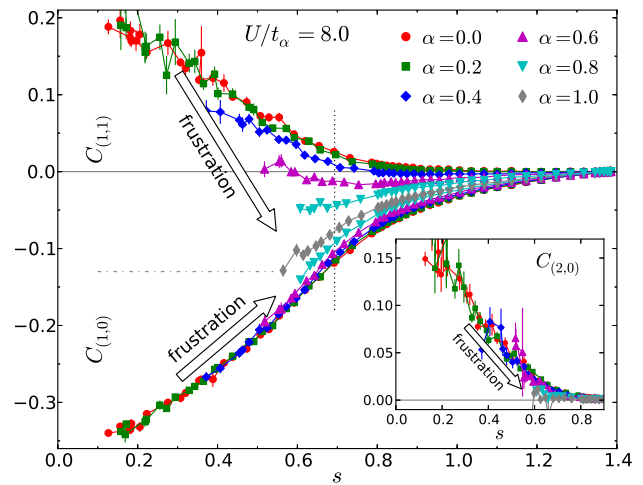}
	\caption{(Color online) Short-range spin correlation functions $C_{(1,0)}$ and $C_{(1,1)}$ at 
	$U = 8 t_\alpha$. The vertical dotted line indicates $s^*\equiv\ln(2)$.
	Inset: spin correlation function $C_{(2,0)}$.}
	\label{fig:corr}
\end{figure}
The strongest sensitivity to $\alpha$ is seen in $C_{(1,1)}$ (upper set of curves in the main panel of \reff{corr}), 
i.e., the spin correlations across the frustrating ``diagonal'' bond. 
This is not surprising, as the direction $(1,1)$ becomes equivalent to $(1,0)$ in the triangular limit 
$\alpha=1$ (grey diamonds in \reff{corr}), where, consequently, $C_{(1,1)}$ must agree with $C_{(1,0)}$. 
More quantitatively, the hopping along the $(1,1)$ diagonal induces an AF superexchange proportional to $\alpha^2$, 
consistent with the AF shifts seen in $C_{(1,1)}$.

In contrast, the lower set of curves in the main panel of \reff{corr}, representing $C_{(1,0)}(s)$, nearly collapse 
in the range $0 \le \alpha\le 0.6$. Even at $\alpha=1.0$,  $|C_{(1,0)}|$ is reduced by only about 
$20\%$ at constant entropy $s$. This has to be contrasted with a reduction by about $60\%$ that is observed at constant 
$T$, as indicated by the upward arrow. Evidently, the cancellation effect 
is more effective for $C_{(1,0)}(s)$ than for $D(s)$ at $U=8\,t_\alpha$
This is even more true for $C_{(2,0)}(s)$, shown in the inset of \reff{corr}, with hardly any effect of $\alpha$ seen for $s\le 0.6$; only the strongest frustration $\alpha=1$ has a significant effect, as it practically eliminates this longer-range correlation in the range $0.6\lesssim s \lesssim 0.7$. 

The cancellation effects clearly expose the relevant physics: a (nearly) 
perfect cancellation of frustration effects in $C_{\mathbf{r}}(s)$ at strong coupling means that the entropy 
$s(T)$ is dominated by spin physics, which is to be expected at $s<s^*$ in this limit.  This is not the case at weak 
coupling, where Fermi liquid physics is equally important, which explains why frustration effects remain so 
pronounced [in $C_{\mathbf{r}}(s)$ and $D(s)$] 
at constant $s$ at weak coupling $U=4\,t_\alpha$ (cf.\ \reff{U4}). 

\begin{figure} 
  \includegraphics[width=\columnwidth]{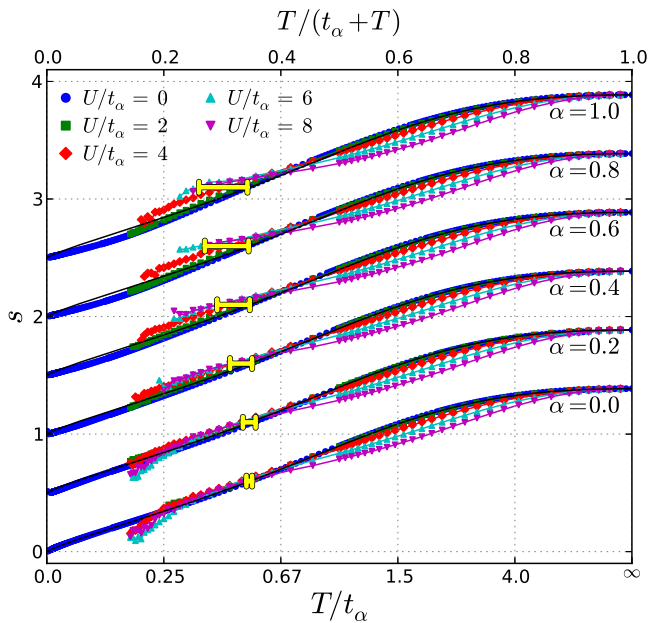}
  \caption{(Color Online) Entropy $s$ versus temperature $T/t_\alpha$ at various interactions $U/t_\alpha$. 
  Data at frustration $\alpha>0$ is offset vertically (by $2.5\, \alpha$); black solid lines represent 
  noninteracting results for the square lattice ($\alpha=0$) as a common reference. Horizontal bars indicate 
  the effect of adiabatic cooling at entropy $s=0.6$, which increases roughly linearly with $\alpha$.
  }  
  \label{fig:entropy}
\end{figure}

\myparagraph{Adiabatic cooling}
Before concluding the paper, let us return to the impact of geometric frustration 
on adiabatic cooling \cite{Werner:2005}. This cooling scheme is based on the thermodynamic relation
$c\,(\partial T/\partial U)_s = T\,(\partial D/\partial T)_U$, where $c$ is the specific heat. $\partial D/dT<0$ 
implies that an adiabatic ramping-up of the interaction between cold fermions in an optical lattice lowers 
their temperature. As depicted in Fig.~\ref{fig:U4} and \ref{fig:D}, we found a wide range of $T$ (or $s$) 
where the slope $\partial D/\partial T$ is negative. To quantify the cooling effect, entropy curves $s(T)$
are plotted in Fig.~\ref{fig:entropy} for a large range of interactions and levels of frustration, with vertical 
offsets proportional to $\alpha$. 

In the charge excitation regime $T\gtrsim 0.7\,t_\alpha$, i.e.,  
$s\gtrsim s^*$, the data at $U>0$ are clearly below the $U=0$ results (blue circles) at all 
$\alpha$. Thus, increasing interactions lead to a reduction of the entropy at constant $T$ or, conversely, 
to a rising temperature at constant $s$. In the unfrustrated case [$\alpha=0$; lowest set of curves in \protect\reff{entropy}], this effect 
disappears at $s<s^*$ and the curves nearly collapse (down to $T/t_\alpha \approx 0.25$ or $s\approx 0.35$). 
In this range, an adiabatic ramp-up of the interaction (from $U=0$ to $U=8\,t_\alpha$) has only a very 
small cooling effect \cite{Paiva:2010}, as indicated by the lower-most horizontal (yellow) bar for $s=0.6$. With 
increasing frustration $\alpha$, the entropy is gradually enhanced by finite interactions, 
leading to an adiabatic cooling. For example, the cooling effect at $s=0.6$ (horizontal yellow bars) is stronger 
by more than an order of magnitude in the triangular limit $\alpha=1$ than on the square lattice ($\alpha=0$).

\myparagraph{Conclusions}
Despite many challenges, rapid progress has been made in the last few years towards the
realization of antiferromagnetic order in optical lattices.  Novel lattice geometries 
are now being explored both for selectively reducing the entropy per particle in parts of the system and 
for exposing or creating magnetic effects that remain visible at elevated average entropy. The
recent detection of AF signatures in tunable dimerized lattices \cite{Greif:2012} is an important 
step in this direction, although the singlet physics of dimers is quite distinct from the long-range 
AF order in higher dimensions.

Our study was aimed at exposing more generic AF physics by selectively suppressing AF order of a bipartite (square) lattice via tunable geometric frustration (by diagonal hopping $t'=\alpha\, t$, as in cold-atom experiments \cite{Greif:2012}) and looking for responses, i.e. sensitivity to $\alpha$, of clearly magnetic character. 
We have demonstrated a clear AF signature, namely an enhancement of the double occupancy $D(s)$ by frustration at weak coupling 
[cf. \reffa{U4}], that extends up to $s\approx \ln(2)$ and should be in reach of cold-atom experiments. 
The approach to this regime could be monitored using the characteristic evolution (and dependence 
on $\alpha$) of spin correlations, which extend to even higher entropies. 

A crucial prerequisite for exposing this physics was our elimination of bandwidth effects by 
scaling the interaction $U$ 
\footnote{In cold-atom experiments, with fixed interaction $U$, $t$ should be scaled with $\alpha$ such that $t_\alpha$ remains constant, within a few percent. Alternatively, one could also introduce a perpendicular hopping $t_z$, between square lattice planes, with $t_z^2 + t'^2 = t^2$ (and $t$ constant) \cite{Bluemer:cubic-tri}.}
with a suitable parameter $t_\alpha$ [cf.\ \refq{scaling}].  
At strong coupling, tunable frustration leads to  AF signatures (at fixed temperatures) which are nearly offset by concurrent entropy changes in curves $D(s)$ [and $C_{\mathbf{r}}(s)$] and, therefore, hardly observable in (adiabatic) cold-atom experiments.
Thus, the optimal interaction $U\approx 4 t_\alpha \lesssim 5 t$ for studying AF physics via tunable geometric frustration is substantially lower than the condition for maximizing the Ne\'el temperature in the cubic case \cite{Kozik:2013}.

\myparagraph{Acknowledgements} 
We thank D. Rost, L.\ Tarruell, and R.R.P. Singh for valuable discussions. RTS was supported under an ARO Award 
with funds from the DARPA OLE Program.  CCC was supported by DOE-DE-FC0206ER25793. NB and EVG gratefully 
acknowledge support by the DFG within the Collaborative Research Centre SFB/TR 49.

\bibliography{reference}

\end{document}


{\center
\section*{Supplemental Material for ``Discriminating antiferromagnetic signatures\\[0.5ex] in ultracold fermions by tunable geometric frustration''}

{\it Chia-Chen Chang, Richard T. Scalettar, Elena V. Gorelik, and Nils Bl\"umer}

\vspace{2ex}
}

\hspace{0.09\textwidth}
\begin{minipage}{0.8\textwidth}
All results presented in the main text are based on determinantal quantum Monte Carlo simulations using a fixed Trotter discretization $\Delta\tau t_\alpha=0.04$ and a fixed cluster size of $8\times 8$. In this supplement, we show that the resulting Trotter error is negligible, compared to statistical errors, and that the finite-size error is small enough not to alter any of our conclusions.
\end{minipage}

\vspace{2ex}
In order to quantify the Trotter errors at weak coupling $U=4\, t_\alpha$, we performed additional simulations using discretizations $\Delta\tau\, t_\alpha \in \{0.05, 0.08, 0.1\}$. As shown in \reff{trotter_ext} for the temperature $T=t_\alpha/4$, the resulting estimates of the double occupancy $D$ (symbols) depend perfectly linearly on $\Delta\tau^2$ for each value of $\alpha$, within error bars, so that a reliable extrapolation to $\Delta\tau$ is possible using least square fits (colored lines). However, already the results at $\Delta\tau t_\alpha=0.04$ (as used in the main text) are converged within the error bars of the individual data points.
\begin{figure}[b]
	\includegraphics[width=0.6\columnwidth]{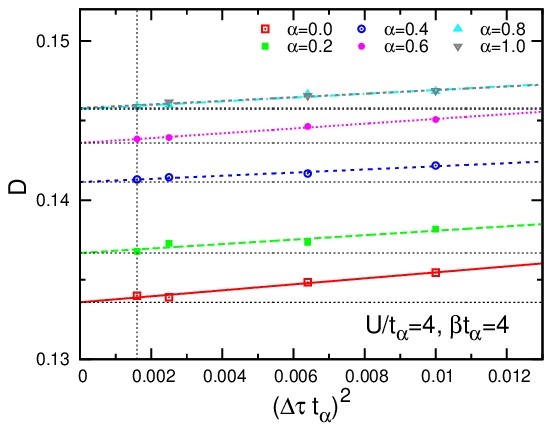}
	\caption{(Color online) The effect of finite Trotter discretization $\Delta\tau$ on DQMC estimates of the double occupancy $D$ at weak coupling $U = 4\, t_\alpha$ and variable frustration $\alpha$. Colored lines show least square fits linear in $(\Delta\tau)^2$. The thin vertical line marks the value of $\Delta\tau t_\alpha=0.04$ used in the main paper. Thin horizontal lines correspond to extrapolated values of $D$.}
	\label{fig:trotter_ext}
\end{figure}

This is seen in \reff{u4_trotter} for the full entropy regime of interest: while the finite-$\Delta \tau$ estimates of $D$ (thin grey symbols) appear slightly shifted upwards, by less than 0.001, in comparison with the extrapolated results (colored symbols) these deviations do not exceed the statistical uncertainties of the individual data points.
Since the deviations are also roughly homogeneous, i.e. independent of $\alpha$ and $s$, they clearly do not affect our conclusions in any way.
\begin{figure}
	\includegraphics[width=0.6\columnwidth]{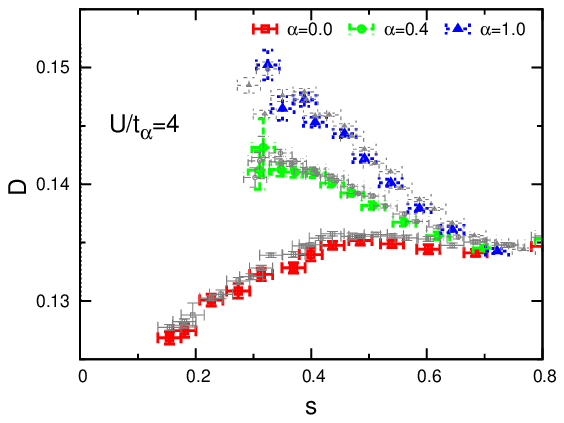}
	\caption{(Color online) The effect of finite Trotter discretization $\Delta\tau$ on DQMC estimates of the double occupancy $D$ at weak coupling $U = 4\, t_\alpha$ and variable frustration $\alpha$. The data at  $\Delta\tau t_\alpha=0.04$ (thin grey symbols) are compared with corresponding $\Delta\tau$-extrapolations (bold colored symbols).}
	\label{fig:u4_trotter}
\end{figure}

At strong coupling $U=12.25\, t_\alpha$, we checked the accuracy of the simulations at  $\Delta\tau\, t_\alpha=0.04$ by  comparisons with previously computed numerically exact data for the square lattice [1]. \reff{u12_trotter} demonstrates that both the double occupancy, and the entropy per particle data used in the main paper match the extrapolated quantities with great accuracy. Thus, Trotter errors appear as irrelevant also at strong coupling.
\begin{figure}
	\includegraphics[width=0.6\columnwidth]{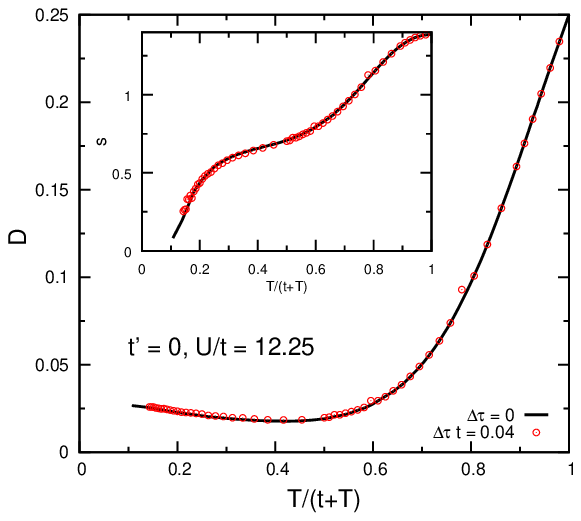}
	\caption{(Color online) The effect of finite Trotter discretization $\Delta\tau$ on DQMC estimates of double occupancy $D$ (main panel) and entropy per particle $s$ (inset) at strong coupling $U = 12.25\, t_\alpha$ for the square lattice ($\alpha=0$). The data for  $\Delta\tau t_\alpha=0.04$ (thin dashed lines) are compared with corresponding $\Delta\tau$-extrapolations [1] (bold solid lines).}
	\label{fig:u12_trotter}
\end{figure}

As all physics is increasingly local at large $U$ (and half filling), the strongest finite-size effects can be expected at weak coupling. Indeed, quite significant deviations are seen for the unfrustrated case ($\alpha=0$) at $U = 4t$ in \reff{FS} between the estimates of spin correlation functions and double occupancy obtained on a $8\times 8$ cluster (indicated by a vertical dotted line) and the thermodynamic limits (horizontal dotted lines) at the lowest temperature $T/t=0.2$. However, this corresponds to an entropy $s=0.25$ which is not of direct experimental interest (yet); in the more relevant entropy range $0.4\lesssim s \lesssim 0.5$ (and above), the finite-size effects at $L=8$ are only slightly larger than the statistical error bars. As larger lattice sizes lead to worse sign problems at significant frustration $\alpha>0$, our choice appears as a good compromise for our comprehensive study of frustration effects.
\begin{figure}
	\includegraphics[width=0.6\columnwidth]{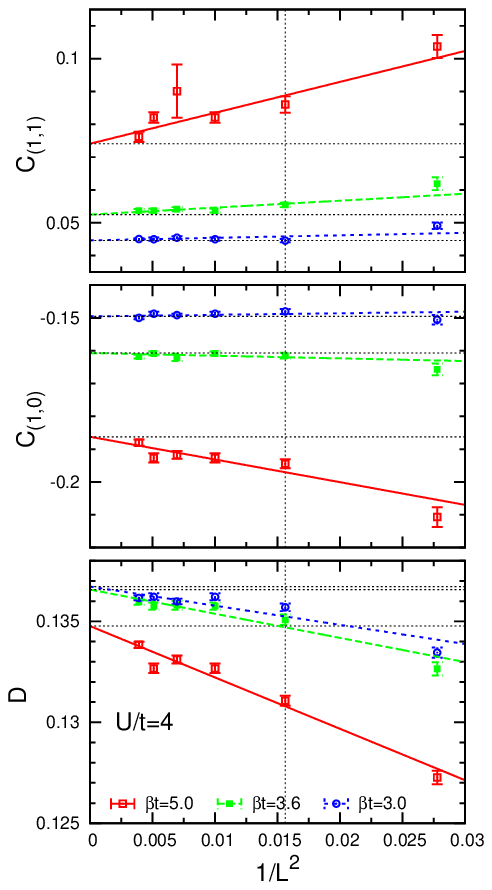}
	\caption{(Color online) Finite-size effects on the DQMC estimates of double occupancy $D$ and spin-spin correlations $C_{(1,0)}$ and $C_{(1,1)}$ at week coupling $U = 4\, t_\alpha$ for square lattice ($\alpha=0$) for a set of thermal parameters: $\beta t=5.0$, $s=0.25\pm 0.02$  (open squares and solid lines), $\beta t=3.6$, $s=0.4\pm 0.02$ (filled squares and dashed lines), $\beta t=3.0$, $s=0.46\pm 0.02$  (open circles and sort-dashed lines). Colored lines show linear least square fits with respect to $L^{-2}$. Thin vertical line marks the value of $L=8$ used in the main paper. Thin horizontal lines correspond to the extrapolated values of observables.}
	\label{fig:FS}
\end{figure}

\vspace{1ex}
\noindent
[1] E. V. Gorelik, D. Rost, T. Paiva, R. Scalettar, A. Kl\"umper, and N. Bl\"umer, Phys. Rev. A {\bf 85}, 061602(R) (2012).